\documentclass[11pt]{article}
\usepackage{graphicx}

\newcommand{\BABARPubYear}    {03}

\newcommand{\BABARConfNumber} {18}
\newcommand{\SLACPubNumber} {10102}
\newcommand{\LANLNumber} {0000}

\input babarsym

\long\def\inst#1{\par\nobreak\kern 4pt\nobreak
    {\it #1}\par\vskip 10pt plus 3pt minus 3pt}

\begin{document}
{\pagestyle{empty}

\begin{flushright}
\babar-CONF-\BABARPubYear/\BABARConfNumber \\
SLAC-PUB-\SLACPubNumber \\
hep-ex/\LANLNumber \\
August 2003 \\
\end{flushright}

\par\vskip 5cm

% Title of the paper
\begin{center}
  \Large \bf \boldmath A Measurement of the Total Width, the
  Electronic Width and the Mass of the $\Upsilon(10580)$ Resonance.
\end{center}
\bigskip
\begin{center}
\large The \babar\ Collaboration\\
\large and\\
\large the PEP-II Machine Group\\
\mbox{ }\\
\today
\end{center}
\bigskip \bigskip
% Abstract
{\abstract We present a preliminary measurement of the resonance
  parameters of the $\Upsilon(10580)$ resonance with the \babar\ 
  detector at the SLAC \pep2\ asymmetric $B$ factory. We measure the
  total width $\Gamma_{\rm tot}$ to be $(20.7\pm1.6\pm2.5) \mev$, the
  partial electronic width $\Gamma_{ee} = (0.321\pm0.017\pm0.029)
  \kev$ and the mass $M = (10.5793\pm0.0004\pm0.0012) \gevcc$. }

\vfill
\begin{center}
  Contributed to the XXI$^{\rm st}$ International Symposium on Lepton
  and Photon Interactions at High~Energies, 8/11 --- 8/16/2003,
  Fermilab, Illinois USA
\end{center}

\vspace{1.0cm}
\begin{center}
{\em Stanford Linear Accelerator Center, Stanford University, 
Stanford, CA 94309} \\ \vspace{0.1cm}\hrule\vspace{0.1cm}
Work supported in part by Department of Energy contract DE-AC03-76SF00515.
\end{center}

\newpage
} % end of pagestyle{empty}

% Input author list file
\begin{center}
\small

The \babar\ Collaboration,
\bigskip

%% author list as of 02-Jun-2003 (595 authors)
%
B.~Aubert,
R.~Barate,
D.~Boutigny,
J.-M.~Gaillard,
A.~Hicheur,
Y.~Karyotakis,
J.~P.~Lees,
P.~Robbe,
V.~Tisserand,
A.~Zghiche
\inst{Laboratoire de Physique des Particules, F-74941 Annecy-le-Vieux, France }
% ...

\end{center}\newpage

% The body of the paper starts here
\section{Introduction}
\label{sec:Introduction}

The $\Upsilon(10580)$ resonance is the lowest lying $b\bar{b}$ state
above the open bottom threshold. Its strong decay into two $B$ mesons
is not suppressed by the OZI rule \cite{OZI}, so that this decay
channel is dominant. The total decay width $\Gamma_{\rm tot}$ of the
$\Upsilon(10580)$ is therefore much larger than the widths of the
lower mass $b\bar{b}$ states, which allows a direct measurement
of $\Gamma_{\rm tot}$ at an $e^+e^-$ collider. 

Although the state has been known for almost 20 years, its mass and width
are still only measured with relatively large uncertainties, and
central values from different experiments show substantial variation
\cite{ARGUS,CLEO1,CLEO2,CUSB}.  We present new measurements of the
mass and total and electronic widths of the $\Upsilon(10580)$ at the
\pep2 storage ring with errors on the mass and total width much lower
than the present world average.

\section{\boldmath The \babar\ experiment}
\label{sec:babar}

The data used in this analysis were collected with the \babar\ 
detector \cite{BabarNim} at the \pep2\ storage ring \cite{PEP2}.  The
data set comprises three energy scans of the $\Upsilon(10580)$ and one
scan of the $\Y3S$ resonance.

The \pep2\ $B$ factory is an asymmetric $e^+e^-$ collider designed to
operate at a luminosity of $3\times 10^{33} \cms $ and a
center-of-mass (CM) energy around $10.58\gev$. The energy of the
electron beam is about $9.0\gev$, that of the positron beam
$3.1\gev$, resulting in a Lorentz boost to the $\Upsilon(10580)$
resonance of $\beta\gamma = 0.56$. The energy of the electron beam was
varied while the positron beam energy was held constant during the
energy scans of the \Y3S and $\Upsilon(10580)$ resonances. The
intrinsic CM energy spread of \pep2\ is about 4.6 MeV.

\babar\ is a solenoidal detector optimized for the asymmetric beam
configuration at \pep2. Charged particle momenta are measured in a
tracking system consisting of a 5-layer, double-sided readout, silicon
vertex tracker (SVT) and a 40-layer drift chamber (DCH) filled with a
mixture of helium and isobutane, both operating in a 1.5 T
superconducting solenoidal magnet. The electromagnetic calorimeter
(EMC) consists of 6580 CsI(Tl) crystals arranged in barrel and forward
endcap subdetectors. Muons and long-lived neutral hadrons are
identified in the instrumented flux return (IFR), composed of
resistive plate chambers and layers of iron. A detector of internally
reflected Cherenkov light (DIRC), together with dE/dx information from
the DCH and SVT, provides separation of kaons and pions.

\section{\boldmath Parametrization of the $\Upsilon(10580)$ resonance shape}
\label{sec:theory}

The $\Upsilon(10580)$ resonance is dominantly a $4S$ state, but its 
shape is modified by interference with the $\Upsilon_1(3D)$ and the
$\Y5S$ states, and by coupled-channel effects at higher energies where
$BB^*$ and $B^* B^*$ production open up.  A full spectroscopic fit of the 
energy region around the $\Upsilon(10580)$ would require both a scan over a much
wider region and more detailed theoretical models, which are both unavailable. 
Fortunately, the $\Upsilon(10580)$ is sufficiently well
isolated to be still treated as a simple resonance, which is therefore
our approach in this analysis.

The $\Upsilon(10580)$ lies only about 20 MeV above the kinematic
threshold for open bottom production, while the total width is
10--20 MeV \cite{PDG2002}.  A model derived from a pure spectroscopic
$4S$ state together with a relativistic Breit-Wigner function is used
to obtain a good approximation of the shape below the opening of new
channels. This analysis is therefore restricted to data points taken
at CM energies below the $BB^*$ threshold at 10.604 GeV. Systematic
effects caused by neglecting interference effects are estimated
by a comparison with a simple non relativistic Breit-Wigner function.

The cross section of the process $\epem \rightarrow \upsbb$ is given as
\begin{equation}
\sigma_0(s) = 12\pi \frac{\Gamma^0_{ee}\Gamma_{\rm tot}(s)}{(s-M^2)^2 + M^2\Gamma_{\rm tot}^2(s)},
\label{purereso}
\end{equation}
where $\Gamma^0_{ee}$ is the partial decay width into \epem,
$\Gamma_{\rm tot}$ is the total decay width and $M$ is the mass of the resonance.
In this equation $\Gamma^0_{ee}$ is taken as a constant and the approximation 
$\Gamma_{\rm tot}(s) \approx \Gamma_{\Y4S\rightarrow B\Bbar}(s)$ is used.

In this analysis the quark pair creation model (QPCM) \cite{QPCM} is
used to describe the energy dependent width $\Gamma_{\rm tot}(s)$. In the
QPCM the decay of the bound $b\bar{b}$ state is described by the
creation of a light quark-antiquark pair from the vacuum. The $b$ and
$\bar{b}$, respectively, each form a $B$ meson with one of the light
quarks. The matrix element for this decay is given by a spin dependent
amplitude and an overlap integral of the $\Y4S$ and $B$-meson wave functions:
\begin{equation}
  I_4(m,{\bf q}) = \int Y^m_1(2{\bf q} - {\bf Q}) \psi_{\rm \Y4S}({\bf Q})\psi_B({\bf Q}-h{\bf q}) 
  \psi_{\rm \bar{B}}(-{\bf Q}+h{\bf q})\, \mathrm{d}^3Q,
\end{equation}
 where {\bf q} is the
momentum of the $B$ meson, {\bf Q} the momentum of the $\Y4S$ and
$h=2m_b/(m_b+m_q)$ with the quark masses $m_b$ and $m_q$. The
spherical tensor $Y^m_1$ represents the wave function of the created
quark-antiquark pair, and $\psi_B$ and $\psi_{\rm \Y4S}$ are the wave
functions of the $B$ and the $\Y4S$ mesons.  The wave
functions $\psi_B$ and $\psi_{\rm \Y4S}$ are approximated by
harmonic oscillator wave functions using the parametrization of the ARGUS
collaboration \cite{ARGUS}.

The resonance shape parametrized by equation (\ref{purereso}) is
significantly modified by QED corrections \cite{QEDcor1,QEDcor2}. The
partial width $\Gamma_{ee}$ 
is related to the lowest order partial width
$\Gamma_{ee}^0$ by $\Gamma_{ee} \approx
\Gamma^0_{ee}(1+\delta_{\rm vac})$, where $\delta_{\rm vac}$ is the vacuum
polarization of the photon propagator \cite{Gamee}. Hence the cross
section including radiative corrections of {\cal O}($\alpha^3$) is
given by
\begin{equation}
\tilde{\sigma}(s) = \int^{\kappa_{\rm max}}_0 12\pi \frac{\Gamma_{ee}\Gamma_{\rm tot}(s')}{(s'-M^2)^2 + M^2\Gamma_{\rm tot}^2} \beta (\kappa^{\beta-1}(1+\delta_{\rm vert}))\, \mathrm{d}\kappa,
\label{correctedform}
\end{equation}
where $\kappa = \frac{2E_{\gamma}}{\sqrt{s}}$ is the energy of the
radiated photon normalized to the CM energy, $\kappa_{\rm max} = 1-\frac{4m_e^2}{s}$, $\beta =
\frac{2\alpha}{\pi}(\ln \frac{s}{m_e^2}-1)$, $\delta_{\rm vert} =
\frac{2\alpha}{\pi}(\frac{3}{4} \ln \frac{s}{m_e^2} -1
+\frac{\pi^2}{6})$ is the vertex correction of the
$e^+e^-\gamma$-vertex and $s' = s(1-\kappa)$.

To obtain the experimentally observed cross section, the cross section
given by equation (\ref{correctedform}) has to be averaged over the
energy distribution of the colliding beams. The CM energies $\sqrt{s'}$
of the actual $e^+e^-$ collisions have a Gaussian distribution about
the mean energy $\sqrt{s}$ with standard deviation $\Delta$. The
experimental cross section is therefore given by
\begin{equation}
\sigma(s) = \int \tilde{\sigma}(s') \frac{1}{\sqrt{2\pi} \Delta} \exp \left( -\frac{(\sqrt{s'}-\sqrt{s})^2}{2\Delta^2}\right) \,\mathrm{d}\sqrt{s'}.
\label{finalformula}
\end{equation}

\section{\boldmath Analysis method}
\label{sec:Analysis}

In order to determine the $\Upsilon(10580)$ resonance parameters, the
energy dependence of the cross section $\sigma_{b\bbar}$ of the
reaction $\epem \rightarrow \Upsilon(10580) \rightarrow B\bar{B}$ has
to be measured in an energy interval around the resonance mass. We
determine the shape of the $\Upsilon(10580)$ resonance from three
short energy scans where the cross section is measured from small data
samples at several CM energies and in the continuum below the
$B\Bbar$ threshold.  This is combined with a precise measurement of
the peak cross section from a high statistics data set taken close to
the peak in the course of $B$-meson data accumulation.  While the scan
data provide information mainly on the width and peak position, the
high precision cross section gives the absolute normalisation with
good accuracy.

The visible hadronic cross section measured from the number of
hadronic events $N_{\rm had}$ and the luminosity $L$ is related to
$\sigma_{b\bbar}$ via
\begin{equation}
\label{form:fit}
\sigma^{\rm vis}(s) \equiv \frac{N_{\rm had}}{L} = \varepsilon_{b\bbar} \sigma_{b\bbar}(s) + \frac{P}{s},
\label{visform}
\end{equation}
where $\varepsilon_{b\bbar}$ is the detection efficiency for \upsbb and
$P$ a constant, which describes the background level of non-$B\Bbar$
events.  A minimum $\chi^2$ fit of (\ref{form:fit}) is done to the
data points ($\sigma^{\rm vis}_{\rm i}$,$\sqrt{s_i}$). The $b\bar{b}$ cross section
is given by the cross section formula (\ref{finalformula}). The
efficiency $\varepsilon_{b\bbar}$, which is considered as energy
independent, is determined for each scan by the peak cross section measured from
the high statistics on-resonance data set.\\
\\
Any selection of hadronic events will have backgrounds from two
classes of sources. Events like $e^+ e^- \rightarrow e^+ e^- (\gamma
\rightarrow) e^+ e^-$ (radiative Bhabhas with converted photon), $e^+
e^- \rightarrow q \qbar (\gamma)$ (continuum) or $e^+ e^- \rightarrow
\tau^+ \tau^- (\gamma)$ contribute to the background. These events all
have cross sections $\sigma \propto 1/s$, at least on a narrow energy
interval, which allows us to describe this background component in a fit
to the data. The second class of background originates from two photon
processes $\gamma \gamma \rightarrow X$ or beam-gas interactions,
which do not scale in a simple way with energy changes, and the latter
even depend on the vacuum in the beam pipe rather than on the beam
energy.  This kind of background cannot be taken into account in the
fit of the resonance. Therefore the event selection has to reduce this
background as much as possible.\\
\\
The selection of hadronic events is based on two main cuts: A high
multiplicity $N_{\rm ch}$ of charged tracks that originate from the beam
crossing region and an event shape cut using the normalized second
Fox-Wolfram moment $R_2$, which is smaller for $B\Bbar$ events than
for background events. Additional cuts as given below are applied to the
different data samples to reduce the beam-gas and $\gamma\gamma$
background.

The luminosity is measured from $e^+e^- \rightarrow \mu^+ \mu^-$
events. These events are selected by requiring a track multiplicity
$N_{\rm ch} \ge 2$ and an invariant mass of a pair of tracks greater
than $7.5\gevcc$. A cut on the cms acolinearity is applied to reject
cosmic rays. At least one of the tracks must have associated energy
deposited in the calorimeter. Bhabha events are vetoed by requiring
that none of the tracks has an associated energy deposited in the
calorimeter of more than 1 GeV.

The energy spread $\Delta$ of the collider has to be known in order to
extract $\Gamma_{\rm tot}$ from the observed resonance shape.  The
energy spread is measured from a scan of the narrow \Y3S resonance.
In addition the $\Y3S$ scan provides a calibration of the product of
the beam energies, which allows a precise measurement of the
$\Upsilon(10580)$ mass.

\subsection{\Y3S scan}
\label{subsec:y3s}

The \Y3S scan consists of 10 data points at different CM energies.
The visible cross section $\sigma^{\rm vis}$ is measured for each
energy. The \Y3S decays dominantly via the three-gluon graph.
Therefore the angular distribution of the decay products of the \Y3S
is more isotropic than the continuum background, which allows us to
select \Y3S events with cuts similar to the $B\Bbar$ selection.  In
particular the cuts $R_2 < 0.4$ and $N_{\rm ch} \ge 3$ are used to
select hadronic events. Additionally the invariant mass of all tracks
is required to be greater than 2.2 \gevcc.

\begin{figure}[!htb]
\begin{center}
  \includegraphics[height=7cm]{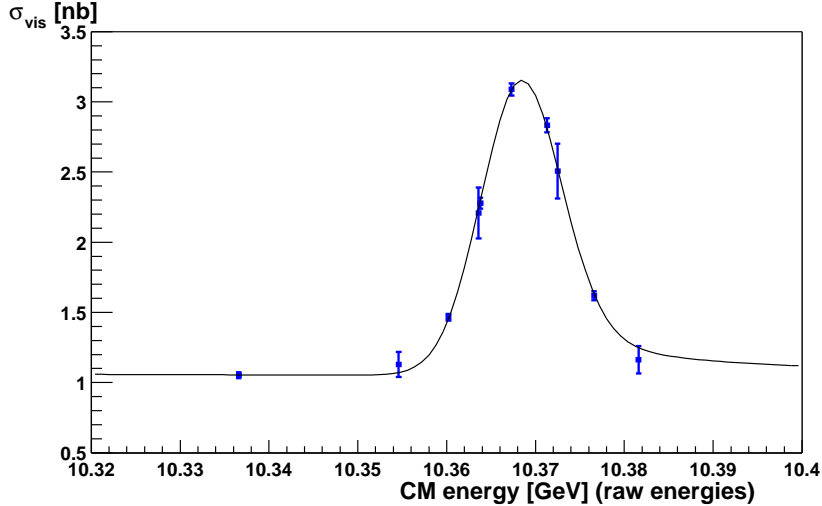}
\caption{Data points and fit of the \Y3S resonance scan.}
\label{fig:y3s}
\end{center}
\end{figure}

The branching fraction of the \Y3S into $\mu^+\mu^-$ is $(1.81\pm0.17)
\%$ \cite{PDG2002}, corresponding to a cross section of roughly $0.1
\nb$ for resonant muon-pair production. Therefore the luminosity is
determined from Bhabha events for the data points of the \Y3S scan.
Figure \ref{fig:y3s} shows the data points and the result of a fit to
these points.

The Breit-Wigner function (\ref{purereso}) of the true \Y3S resonance
is approximated by a delta function, because the width of the \Y3S,
$\Gamma^{\rm 3S}_{\rm tot} = (26.3\pm3.5)\kev$ \cite{PDG2002}, is very small
compared to the energy spread of \pep2 of about $4.6 \mev$. Hence the cross
section including radiative corrections and averaging over the energy
spread $\Delta$ is given by
\begin{equation}
\sigma(s) = \frac{6\pi^2}{M^2\Delta\sqrt{2\pi}} \frac{\Gamma_{ee}\Gamma_{\rm had}}{\Gamma_{\rm tot}}\beta
\left(\frac{2\Delta}{\sqrt{s}}\right)^{\beta} \cdot \int_0^\infty x^{\beta-1} \exp\left(-\frac{(z-x)^2}{2}\right)\,\mathrm{d}x,
\end{equation}
where $z = \frac{\sqrt{s}-M}{\Delta}$. This cross section is related to the visible cross
section via equation (\ref{visform}).

The visible cross section of equation (\ref{visform}) is fitted to
the data points.  The free parameters of the fit are the energy
spread $\Delta$, the \Y3S mass $M_{3S}^{fit}$, the parameter $P$
describing the background and the efficiency $\varepsilon$
  for selecting \Y3S decays. The ratio $\Gamma_{ee}\Gamma_{\rm
    had}\over\Gamma_{\rm tot}$ is set to $0.45\kev$ \cite{CLEO3S}.
  The result of the fit including the statistical errors is shown in
  the first two rows of table~\ref{tab:y3s}.

\begin{table}[!htb]
\caption{Energy spread $\Delta$ of \pep2 and \Y3S mass at the \pep2 energy scale obtained from a fit to the \Y3S scan data including their statistical errors. The value of the energy spread extrapolated to $10.58\gev$ and the difference of the fitted \Y3S mass to the PDG value are shown as well. The statistical and systematic errors of the latter two quantities are added in quadrature.}
\begin{center}
\begin{tabular}{|l|c|}
\hline
$\Delta$ & $(4.44\pm0.09)\mev$ \\
\hline
$M_{\rm 3S}^{\rm fit}$\vrule width 0pt height 2.3ex& $(10.36798 \pm 0.00009)\gevcc$ \\
\hline
$\Delta$ extrapolated to $\sqrt s=10.58\gev$\vrule  width 0pt height 2.5ex & $(4.63\pm0.15)\mev$ \\
\hline
$M_{\rm 3S}^{\rm PDG} - M_{\rm 3S}^{\rm fit}$\vrule width 0pt height 2.3ex & $(12.8 \pm 0.5)\mevcc$ \\
\hline
\end{tabular}
\end{center}
\label{tab:y3s}
\end{table}      
 
Sources of systematic uncertainties to the fit results are potential
fluctuations of the detector and trigger performance during the $\Y3S$
scan and the precision ($\pm 0.2\mev$) of the determination of the
energy differences between the scan points. In total the systematic
uncertainty is estimated to be $0.11 \mev$ and $0.14 \mevcc$ for the
energy spread and $\Y3S$ mass, respectively.

A comparison of the fitted \Y3S mass $M_{\rm 3S}^{\rm fit}$ with the
world average of $(10.3552\pm0.0005)$ \gevcc \cite{vepp4} provides a
calibration of the product of PEP-II beam energies. The information
obtained from the \Y3S scan that is essential for the measurement of
the $\Upsilon(10580)$ parameters is shown in table~\ref{tab:y3s}. The
energy spread measured at the \Y3S is extrapolated to $10.58\gevcc$ by
scaling the spread of the high energy beam with the square of its
energy. An extrapolation of the spread of the low energy ring is not
necessary, because its energy was always held constant. The
extrapolation results in a spread of $4.63\mev$. The energy spread
during two of the three $\Upsilon(10580)$ scans was $0.2\mev$ larger
due to a different magnet configuration of \pep2.

\subsection{Measurement of the peak cross section}
\label{subsec:peakXS}

The $b\bar{b}$ cross section at the peak of the $\Upsilon(10580)$
resonance is determined from the energy dependence of
$\sigma_{b\bar{b}}$ measured from a high statistics data set. The
cross section $\sigma_{b\bbar}$ is given by

\begin{equation}
  \sigma_{b\bbar} = \frac{N_{\rm had} - N_{\rm \mu\mu} \cdot R_{\rm off} \cdot r}{\varepsilon_{b\bbar}L},
\end{equation}
where $N_{\rm \mu\mu}$ is the number of muon pairs, $R_{\rm off}$ the
ratio of hadronic events to muon pairs below the resonance and $r\approx 1$ a
factor estimated from Monte Carlo events to correct variations of
cross sections and efficiencies with the CM energy.

\begin{figure}[!h]
\begin{center}
\includegraphics[height=6.5cm]{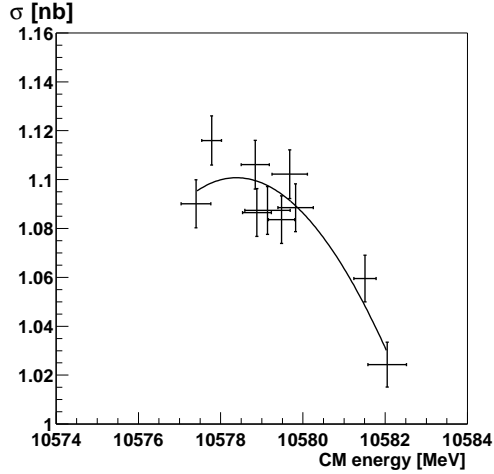}
\caption{$\sigma_{b\bbar}$ vs. CM energy.}
\label{fig:peakXS}
\end{center}
\end{figure}

A track multiplicity of $N_{\rm ch} \ge 3$ and a topology cut of $R_2
< 0.5$ is applied to select these hadronic events. Events from $\gamma
\gamma$ interactions and beam-gas background are reduced by selecting
only events with a total energy greater than 4.5 GeV. Beam-gas
interactions are additionally reduced by requiring that the primary
vertex of these events lies in the beam collision region.
Figure~\ref{fig:peakXS} shows the cross section at several CM
energies.

A fit of a 3rd order polynomial to the data results in a peak cross
section of $(1.101\pm0.005) \nb$. The systematic error of the peak
cross section is 2\%, which is dominated by uncertainties of the
efficiency $\varepsilon_{b\bbar}$ and the luminosity
determination.  

\subsection{Fit of the $\Upsilon(10580)$ resonance}
\label{subsec:y4sfit}
  
\begin{figure}[!htb]
\begin{center}
\includegraphics[height=7cm]{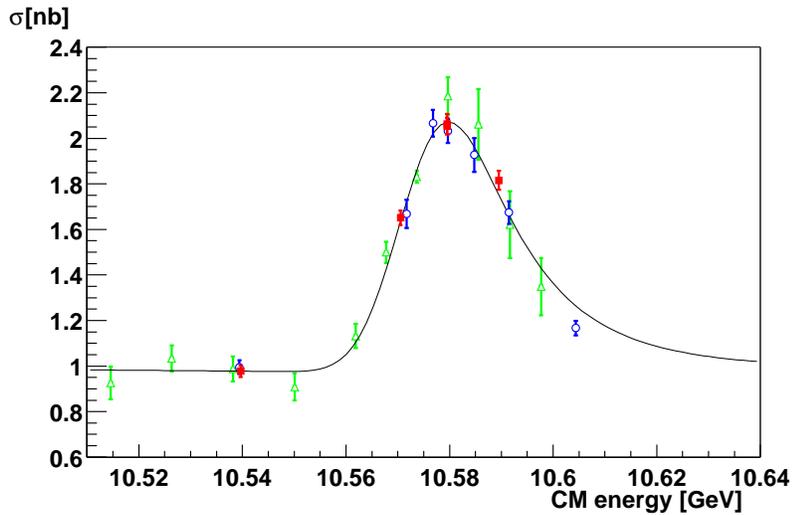}
\caption{Data points and fit of the $\Upsilon(10580)$ resonance scan. Different symbols represent different scans. The small efficiency variations between the different scans are corrected.}
\label{fig:y4sfit}
\end{center}
\end{figure}

The $\Upsilon(10580)$ scan comprises three scans around the resonance
mass. Hadronic events are selected by requiring
$N_{\rm ch} \ge 4$ and $R_2 < 0.3$. The background from beam-gas and
$\gamma\gamma$ interactions is reduced by the cut
$E_{\rm tot}-|P_z| > 0.2 \sqrt{s} $, where $E_{\rm tot}$ is the total
CM energy calculated from all charged tracks and $P_z$ is the
component of the total momentum of all charged tracks along the beam
axis. The data points, ($\sigma^{\rm vis}_i$,$\sqrt{s_i}$), are shown in
figure~\ref{fig:y4sfit}.

The visible cross section of equation (\ref{visform}), with the
$b\bar{b}$ cross section given by (\ref{finalformula}), is fitted to
the data points.  The CM energies of the $\Upsilon(10580)$ scans are
corrected using the shift obtained from the \Y3S fit. The free
parameters are the total width $\Gamma_{\rm tot}$, the electronic
width $\Gamma_{ee}$, the mass $M$ of the $\Upsilon(10580)$, the
background parameter $P$ and the efficiency $\varepsilon_{b\bar{b}}$.
The energy spread of the collider is fixed to 4.63 MeV.  The result
of the fit is shown in table~\ref{tab:fitresults}.  Alternatively the
electronic branching fraction
$$
B_{ee} = {\Gamma_{ee} \over \Gamma_{\rm tot}}
$$
can be
used as a free fit parameter instead of $\Gamma_{ee}$ or
$\Gamma_{\rm tot}$. The fit results of these three quantities are
highly correlated as can be seen from table~\ref{tab:corelmatrix}. 
We therefore quote the results of all three parameters.

\begin{table}[!h]
\caption{$\Upsilon(10580)$ resonance parameters and their statistical errors obtained from a fit to the scan data.}
\begin{center}
\begin{tabular}{|c|c|}
\hline
$\Gamma_{\rm tot}$ &  $(20.7 \pm 1.6)\mev$  \\
\hline
$\Gamma_{ee}$ & $(0.321\pm0.017)\kev$ \\
\hline
$B_{ee} $ & $(1.55\pm0.04)\cdot 10^{-5}$ \\
\hline
$M$ & $(10.5793\pm0.0004)\gevcc$ \\
\hline
$\chi^2$/DoF & 18.3/14 \\
\hline
\end{tabular}
\end{center}
\label{tab:fitresults}
\end{table}

\begin{table}[!h]
\caption{Correlation coefficients of the fit to the $\Upsilon(10580)$ scans. Any combination of two of the 
 three parameters $\Gamma_{\rm tot}$, $\Gamma_{ee}$ and $B_{ee}$ can be used as free parameters in
 the fit.}
\begin{center}
\begin{tabular}{|c|ccc|}
\hline
 & $\Gamma_{ee}$ & $B_{ee}$ & $M$  \\
\hline
$\Gamma_{\rm tot}$ & 0.996 & -0.980 & 0.206 \\
$\Gamma_{ee}$ &  & -0.961 & 0.186 \\
$B_{ee} $ & & & -0.226 \\
\hline
\end{tabular}
\end{center}
\label{tab:corelmatrix}
\end{table}

\section{\boldmath Systematic studies}
\label{sec:Systematics}

The systematic error induced by the assumptions in the quark pair
creation model, e.g., neglecting interference effects with other
spectroscopic states, is estimated by fitting a non relativistic
Breit-Wigner function with a constant decay width $\Gamma_{\rm tot}$
to the scan data.  Half of the deviation between the results of this
fit and the fit using the QPCM is taken as an estimation of the
systematic uncertainty due to the resonance parametrization. The
systematic errors assigned to each parameter are listed in
table~\ref{tab:sys}.

A systematic bias on the fit results could be caused by detector
instabilities or a incorrect energy measurement during a scan. This effect
is estimated by excluding single data points from the fit. The maximum
shift for each fit parameter is taken as a systematic error.
 
The \Y3S scan and the $\Upsilon(10580)$ scans were spread over a
period of three years.  A systematic error of 1.0 MeV is
assigned to the mass measurement due to drifts in the beam energy
determination between the $\Upsilon(10580)$ scans and the \Y3S scan
that are not reflected in the beam energy corrections. Another
contribution to the uncertainty of the mass measurement is caused by
the precision of the \Y3S mass knowledge. The systematic error caused
by the uncertainty of the energy spread of the collider is estimated
by varying the energy spread used in the fit procedure for all three
$\Upsilon(10580)$ scans by its uncertainty of $\pm0.15\mev$. Long term
fluctuation of the energy spread are taken into account by varying
the energy spread of single scans in the fit by $\pm 0.1\mev$ from its
nominal value. The quadratic sum of both contributions is listed in
table \ref{tab:sys}.  In addition the systematic error due to the
uncertainty of the peak cross section is included. The systematic
uncertainties caused by potential energy dependences of the event
selection efficiencies are found to be negligible.

\begin{table}[!htb]
\caption{Summary of systematic uncertainties}
\begin{center}
\begin{tabular}{|l|crc|clc|c|crc|}
\hline
 & \multicolumn{3}{|c|}{$\delta\Gamma_{\rm tot}$ [MeV]} & \multicolumn{3}{|c|}{$\delta \Gamma_{ee}$ [keV]} & $\delta B_{ee}/10^{-5}$ & \multicolumn{3}{|c|}{$\delta M$ [\mevcc]} \\
\hline
 model uncertainty & & 1.4 & & & 0.017 & & 0.03 & & 0.1 & \\
\hline
 systematic bias by single data point & & 2.0 & & & 0.022 & & 0.04 & & 0.3 & \\
\hline
 uncertainty of energy spread & & 0.4 & & & 0.0019 & & 0.03 & & $<$ 0.1 & \\
\hline
 uncertainty of peak cross section & & $<$ 0.1 & & & 0.006 & & 0.03 & & $<$ 0.1 & \\
\hline
 long term drift of energy scale & \multicolumn{3}{|c|}{-} &  \multicolumn{3}{|c|}{-} & - & & 1.0 & \\
\hline
 error on $M_{\rm \Y3S}$ &  \multicolumn{3}{|c|}{-}  &  \multicolumn{3}{|c|}{-}  & - & & 0.5 & \\ 
\hline
\hline
 total error & & 2.5 && & 0.029 & & 0.07 & & 1.2 & \\
\hline
\end{tabular}
\end{center}
\label{tab:sys}
\end{table}

\section{\boldmath Results}
\label{sec:Physics}

In summary we have preliminarily measured the total decay width $\Gamma_{\rm tot}$, the
partial decay width into electrons $\Gamma_{ee}$ and the mass of the
$\Upsilon(10580)$ resonance and find
\begin{eqnarray*}
\Gamma_{\rm tot} & = & (20.7\pm1.6\pm2.5)\mev, \\
\Gamma_{ee} & = & (0.321\pm0.017\pm0.029)\kev, \\
B_{ee} & = & (1.55\pm0.04\pm0.07)\cdot 10^{-5}, \\
M & = &(10.5793\pm0.0004\pm0.0012)\gevcc. \\
\end{eqnarray*}
The measurement of the total width and mass are an improvement in
precision compared to the present world average. 

\section{ACKNOWLEDGMENTS}
\label{sec:Acknowledgments}

We are grateful for the 
extraordinary contributions of our \pep2\ colleagues in
achieving the excellent luminosity and machine conditions
that have made this work possible.
The success of this project also relies critically on the 
expertise and dedication of the computing organizations that 
support \babar.
The collaborating institutions wish to thank 
SLAC for its support and the kind hospitality extended to them. 
This work is supported by the
US Department of Energy
and National Science Foundation, the
Natural Sciences and Engineering Research Council (Canada),
Institute of High Energy Physics (China), the
Commissariat \`a l'Energie Atomique and
Institut National de Physique Nucl\'eaire et de Physique des Particules
(France), the
Bundesministerium f\"ur Bildung und Forschung and
Deutsche Forschungsgemeinschaft
(Germany), the
Istituto Nazionale di Fisica Nucleare (Italy),
the Foundation for Fundamental Research on Matter (The Netherlands),
the Research Council of Norway, the
Ministry of Science and Technology of the Russian Federation, and the
Particle Physics and Astronomy Research Council (United Kingdom). 
Individuals have received support from 
the A. P. Sloan Foundation, 
the Research Corporation,
and the Alexander von Humboldt Foundation.

\end{document}